\documentclass[letterpaper, 10 pt, conference]{ieeeconf}  

\IEEEoverridecommandlockouts                              

\usepackage{multirow}
\usepackage{multicol}
\usepackage{booktabs}
\usepackage{amsmath}
\usepackage{amssymb}
\usepackage{svg}
\usepackage{comment}
\usepackage{subfigure}
\usepackage{subcaption}
\usepackage{epstopdf}
\usepackage{comment}
\usepackage{bm}
\usepackage{booktabs}

\title{\LARGE \bf

Quantum game models for interaction-aware decision-making in automated driving

}

\author{Karim Essalmi$^{1}$$^{,2}$, Fernando Garrido$^{1}$$^{,2}$, and Fawzi Nashashibi$^{1}$
\thanks{$^{1}$Inria Paris, 48 rue Barrault, 75013 Paris, France {\tt\small \{karim.essalmi, fernando.garrido-carpio, fawzi.nashashibi\}@inria.fr}}
\thanks{$^{2}$Valeo Mobility Tech Center, 6 rue Daniel Costantini, 94000 Créteil, France {\tt\small \{karim.essalmi, fernando.garrido\}@valeo.com}}}

\begin{document}

\maketitle
\thispagestyle{empty}
\pagestyle{empty}

\begin{abstract}
Decision-making in automated driving must consider interactions with surrounding agents to be effective. However, traditional methods often neglect or oversimplify these interactions because they are difficult to model and solve, which can lead to overly conservative behavior of the ego vehicle. To address this gap, we propose two quantum game models, QG-U1 (Quantum Game - Unitary 1) and QG-G4 (Quantum Game - Gates 4), for interaction-aware decision-making. These models extend classical game theory by incorporating principles of quantum mechanics, such as superposition, interference, and entanglement. Specifically, QG-U1 and QG-G4 are designed for two-player games with two strategies per player and can be executed in real time on a standard computer without requiring quantum hardware. We evaluate both models in merging and roundabout scenarios and compare them with classical game-theoretic methods and baseline approaches (IDM, MOBIL, and a utility-based technique). Results show that QG-G4 achieves lower collision rates and higher success rates compared to baseline methods, while both quantum models yield higher expected payoffs than classical game approaches under certain parameter settings.
\end{abstract}

\section{Introduction}
Despite recent advancements in the field \cite{advancements}, the deployment of autonomous vehicles in daily life remains a challenging task, primarily due to the complex interactions between road users. Human driver interactions seem to be innate, making it difficult to model these decisions computationally. Nevertheless, incorporating these interactions into decision-making algorithms is crucial for the successful integration of autonomous vehicles into everyday life.

Nowadays, most decision-making algorithms consider interactions in only one direction. This means they assume that other agents interacting with the ego vehicle (EV) will not adapt their behavior in response to the EV's decisions, which simplifies the problem. For example, the studies presented in \cite{cor-mp, cor-mcts} follow this unilateral assumption, which often leads to overly conservative behavior in certain situations.
However, in real-world driving, human drivers consider interactions in a multidirectional way: the decisions of each agent influence and are influenced by others. To accurately model such behavior, decision-making algorithms must be interaction-aware, accounting for these multidirectional influences. 

One approach to accounting for these interactions is Classical Game Theory (CGT). Game theory studies decision-making among rational competing agents in conflict situations. It was initially formalized in the mid-20th century by Von Neumann and Morgenstern in \textit{The Theory of Games and Economic Behavior} \cite{von2007theory} within the field of economics. Since then, game theory has evolved and has been adopted to various fields, including computer science, economics, and psychology \cite{che2024game}.

Even though CGT makes the model interaction-aware, it still suffers from several limitations. The primary one is the assumption that all players in the game are rational, meaning each player seeks to maximize their own payoff. However, in many real-world situations, human decisions are often irrational \cite{songvehicle}.
Additionally, CGT does not adequately account for uncertainties, including epistemic and aleatoric, which are crucial factors in Automated Driving (AD). Another limitation is that, in some cases, the game outcome becomes predictable because CGT relies mainly on payoffs to make decisions. 

To address these limitations, we explore a subfield of Classical Game Theory known as Quantum Game Theory (QGT), which integrates principles of both CGT and quantum mechanics. By leveraging quantum effects such as interference, superposition, and entanglement, QGT significantly expands the strategy space compared to CGT. This expansion leads to new possible outcomes and the emergence of additional equilibria. However, the increased strategy space also makes it more challenging to determine the optimal strategy for a player \cite{silva2022learning}. Additionally, QGT can account for human irrationality, as demonstrated in \cite{zhang2021subjective}. Despite its potential, research on QGT remains largely theoretical and has yet to be widely applied in simulation environments or real-world scenarios. To the best of our knowledge, only three studies in the literature apply quantum concepts to the domain of automated driving \cite{songvehicle, songpedestrian, sinha2025nav}, yet none explore their application to decision-making within this context.

Given these gaps and the existing challenges in deploying self-driving vehicles, we find it valuable to explore the application of QGT for decision-making in AD. In this paper, we introduce QG-U1 (Quantum Game - Unitary 1) and QG-G4 (Quantum Game - Gates 4), two quantum game models designed for the tactical level of the decision-making pyramid \cite{garrido2022review}, also referred to as maneuver planning. Notably, both models can be executed on a classical computer without requiring quantum hardware. Basically, QG-U1 and QG-G4 aim to solve two-player games in which each player has two available strategies.

The main contributions of this paper are:
\begin{itemize}
    \item The formulation of two novel quantum game-theoretic models (QG-U1 and QG-G4) to address the challenges of interaction-aware decision-making.
    \item The application and analysis of these models in real automated driving scenarios involving strategic dilemmas.
    \item A comparative evaluation highlighting the differences between Quantum Game Theory, Classical Game Theory, and baseline decision-making methods in this context.
\end{itemize}

This paper is structured as follows: Section \ref{sec:relatedwork} reviews related work in both classical and quantum game theory. Section \ref{sec::method} presents our approach. Results are then discussed in Section \ref{sec::results}. Finally, Section \ref{sec::conclusion} provides the conclusion and outlines future research directions.

\section{Related Work} \label{sec:relatedwork}
\subsection{Classical Game Theory}
Game theory is an intuitive yet powerful framework for modeling and solving decision-making problems involving high interaction between agents. It has gained significant attention in various domains, including economics, psychology, and computer science \cite{che2024game}. Even in the field of decision-making for automated driving, game theory has established itself as a valuable tool. Since its inception, it has evolved significantly, leading to the development of various models tailored to different problem types, such as cooperative games, extensive games, static games, etc, \cite{ibrahim2021comprehensive}.

\begin{figure}
    \centering
    \includegraphics[width=\linewidth]{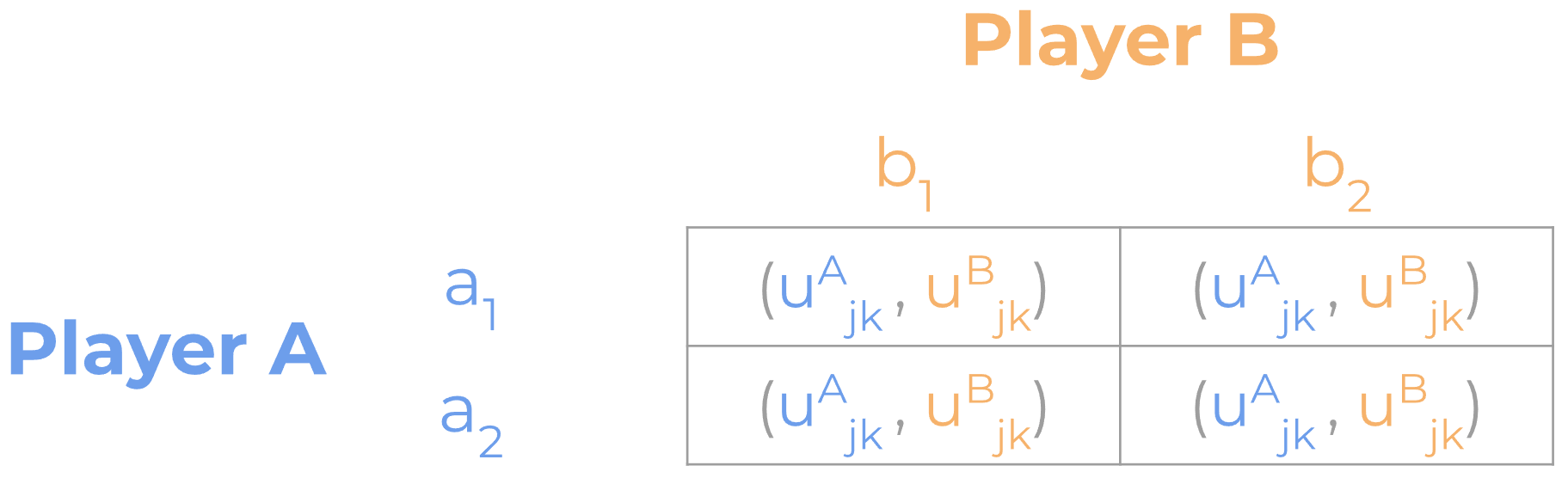}
    \caption{Example of a payoff matrix with two players, A and B, each having two pure strategies: \(\mathcal{A}_A=(a_1,a_2)\) and \(\mathcal{A}_B=(b_1,b_2)\). \(u^i_{jk}\) represents the payoff of player \(i\) when player A chooses strategy \(a_j\) and player B chooses strategy \(b_k\).}
    \label{fig::payoff_table}
\end{figure}

The most common approach to modeling strategic interactions in game theory is through a normal-form game, which defines the number of players, their available strategies, and a payoff matrix that specifies the outcomes for each strategy combination. An example of such a matrix is illustrated in Figure \ref{fig::payoff_table}. Previous works, such as \cite{naidja2024gtp, liu2022three}, model decision-making as a normal-form game and solve it by finding the Nash Equilibrium (N.E.). Where a N.E. is a state in the game where no player can improve their payoff by unilaterally changing their strategy, assuming all other players keep their strategies unchanged. Specifically, in \cite{naidja2024gtp}, authors apply game theory to solve left-turn situations at intersections involving two agents, determining the optimal strategy based on N.E.. Finding the N.E. offers an intuitive and elegant way to solve a game, as each player is satisfied with the outcome. However, in some games, multiple N.E.s may exist, introducing dilemmas that make it difficult to resolve the game solely by identifying an equilibrium.



An alternative approach for solving a game is to model it using the Stackelberg framework, which involves two players: a leader and a follower. The leader first decides on its behavior, and the follower then optimizes its response based on the leader's decision. For example, \cite{burger2022interaction} combines the Stackelberg model with Model Predictive Control (MPC) to enhance decision-making. 

Another elegant technique for solving a normal-form game is level-k reasoning, as used in \cite{garzon2019game, yuan2023scalable}. This method models agent interactions under the assumption that a level-k player makes optimal decisions while considering other agents as level-(k-1) players. By structuring reasoning in this way, level-k reasoning simplifies the process of finding an equilibrium in the game. 

In multi-agent settings, one alternative within game theory is to model their interactions through coalition formation. Instead of competing individually, players form coalitions to compete against other groups of players. In \cite{heshami2024towards}, lane-changing scenarios are modeled as a coalitional game, where the solution is obtained by identifying the Pareto-optimal coalition. 

While driving, uncertainties are often present and can be categorized as aleatoric (due to inherently random effects) or epistemic (caused by a lack of knowledge about the environment). A Bayesian game can account for certain types of uncertainties by modeling a game with incomplete information. The authors of \cite{huang2024integrated} employed a Harsanyi game, which is a type of Bayesian game, to address decision-making in intersections and merging situations. The solution to the game is obtained using what they call Bayesian Coarse Correlated Equilibrium, a generalization of the Bayesian Nash Equilibrium. 

Finally, as learning techniques have become a prominent research direction in recent studies, some works have combined them with game theory, as seen in \cite{zhou2024game, yuan2021deep}, where both studies employed level-k reasoning. Specifically, \cite{yuan2021deep} integrates it with Deep Reinforcement Learning (DRL) in a four-agent intersection scenario, while \cite{zhou2024game} combines it with Q-Learning to approximate the optimal policy based on game outcomes. 

\subsection{Quantum Game Theory}

As mentioned in \cite{pricequantum, wiki_quantum_game_theory}, the differences between Quantum Game Theory and Classical Game Theory are as follows:

\begin{itemize}
    \item \textbf{Initial states can be superimposed}: In CGT, initial states are well-defined and deterministic. For example, in the Prisoner’s Dilemma, players begin with a clear choice: cooperate or defect. In contrast, QGT allows the initial states to exist in a superposition, meaning the system can be in multiple states simultaneously until a measurement collapses it into a definite state.
    
    \item \textbf{Initial states can be entangled}: In CGT, players’ choices are independent of each other. For example, in the Prisoner’s Dilemma, a player's decision to cooperate or defect does not directly influence the other player's decision. As opposed to QGT, where it allows initial to be entangled. Entanglement is a quantum phenomenon where the states of players are correlated in such a way that one player's state cannot be described independently of the others. In fact, the degree of entanglement determines "how quantum" the game is.

    \item \textbf{Superposition of strategies}: In CGT, players select a single strategy (e.g., cooperate or defect) and apply it to the game. The outcome is determined by the combination of the chosen strategies. In QGT, players can utilize a superposition of strategies. Instead of committing to a single strategy, they can apply a quantum operation (represented by a unitary matrix or a quantum gate) that transforms the initial state into a superposition of multiple possible outcomes.
\end{itemize}

Esiert et al. were the first to propose combining game theory with quantum mechanics principles. In their study \cite{eisert1999quantum}, they introduced the Eisert-Wilkens-Lewenstein (EWL) protocol and applied it to the well-known Prisoner's Dilemma. In the classical version of the game, both rational players typically choose to "Defect", as it is the dominant strategy and the Nash Equilibrium. However, when the EWL formalism is applied, the dominant strategy is no longer "Defect". The study also highlights that in an unfair game, where one player uses quantum strategies while the other uses classical strategies, the quantum player can achieve a higher expected payoff. This work marked the first demonstration of the potential of quantum game theory. Since then, several studies have expanded on this idea. For example, in \cite{flitney2002introduction}, the authors explore the quantum version of the Penny Flip game, showing that when Alice plays quantum strategies and Bob uses classical strategies, Alice wins 100\% of the time. In contrast, when both players use classical strategies, each has a 50\% chance of winning. The study also illustrates how QGT can be extended to multi-player games. More recently, \cite{khan2025quantum} demonstrated potential quantum advantages in applying QGT in a trading application. 

As mentioned in the introduction, we found only three studies in the literature that have applied quantum aspects to the domain of automated driving \cite{songpedestrian, songvehicle, sinha2025nav}. 
In \cite{songvehicle}, the authors take inspiration from the study presented in \cite{pothos2009quantum} and adapt the model to the context of AD. Specifically, they focus on highly interactive scenarios involving two vehicles: one attempting to change into the other's lane while the second vehicle simultaneously tries to enter the first vehicle's current lane. The study compares quantum probabilities with classical probabilities and finds that QGT can achieve a higher success rate in predicting the behavior of both vehicles. 

In \cite{songpedestrian}, the authors address the problem of pedestrian trajectory prediction at crosswalks, particularly focusing on whether a pedestrian decides to cross or not. To model this, they combine a classical Bayesian Network with quantum probabilities, introducing the Quantum-like Bayesian (QLB) model. Their approach is then compared with classical methods, highlighting the potential benefits of incorporating quantum probabilities. 

Finally, \cite{sinha2025nav} investigates the application of quantum concepts in the context of automated driving. Although the authors do not adopt the Quantum Game Theory framework, they apply Quantum Reinforcement Learning (QRL) for end-to-end learning, taking raw images as inputs and providing speed and steering angle as outputs. QRL combines quantum computing with reinforcement learning, leveraging the advantages of both. In this work, the authors demonstrate that QRL improves training performance by enabling faster convergence compared to the classical reinforcement learning technique. 

As shown in this subsection, only a few studies have explored Quantum Game Theory, and even fewer have applied it within the context of automated driving. Among these, none of them applied it to decision-making for AD. Moreover, most existing research on QGT remains theoretical and has not yet been implemented in simulation environments or real-world scenarios. This highlights the need for further investigation into the practical applications of QGT. 

\section{Method} \label{sec::method}
\begin{figure}
    \centering
    \includegraphics[width=\linewidth]{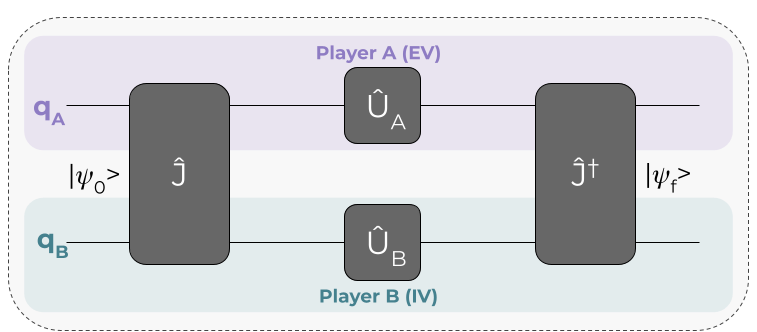}
    \caption{QG-U1 model - Quantum Circuit}
    \label{fig::QG-U1}
\end{figure}

\begin{figure}
    \centering
    \includegraphics[width=\linewidth]{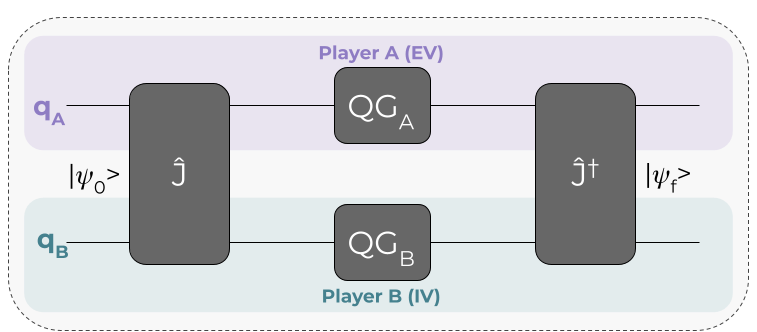}
    \caption{QG-G4 model - Quantum Circuit}
    \label{fig::QG-G4}
\end{figure}

Our approach is inspired by the Eisert-Wilkens-Lewenstein model \cite{eisert1999quantum}, which we adapt to the context of decision-making in automated driving (implemented and simulated on a classical computer, without requiring quantum hardware). Figures \ref{fig::QG-U1} and \ref{fig::QG-G4} illustrate the quantum circuits of our proposed models designed to solve a two-player game with two strategies per player. The core of our approach is defined similarly to a finite, n-person normal-form game \(\mathcal{G} = \Bigl( \mathcal{N}, \mathcal{A}, u \Bigr) \). Where:

\begin{itemize}
    \item \(\mathcal{N} = \{1, \dots, n\}\) is a finite set of players, indexed by \(i\),
    \item \(\mathcal{A}= \mathcal{A}_1 \times \dots \times \mathcal{A}_n\) is the set of all action profiles, where an action profile \(a=(a_1, \dots, a_n)\) represents the actions taken by all players, with \(a_i \in \mathcal{A}_i\) for each player \(i\),
    \item \(\mathcal{A}_i\) is the action set (or pure strategy set) for player \(i\),
    \item \(u=(u_1, \dots, u_n)\) is the profile of utility functions for all players,
    \item \(u_i:\mathcal{A} \rightarrow \mathbb{R}\) is the utility (payoff) function for player \(i\).
\end{itemize}

As stated above, a quantum game differs from a classical game in three main aspects: superposition of the initial states, entanglement, and superposition of strategies \cite{pricequantum}. To account for these differences, in our approach, each player's strategy is defined using qubits. Basically, each player is associated with a qubit that encodes their state in the game.

A qubit \(q_i\) is defined as follows:
\begin{equation} \label{eq:qubit}
    q_i= \alpha_i \big|0 \rangle + \beta_i \big| 1 \rangle
\end{equation}

With \(\alpha_i, \beta_i \in \mathbb{C}\), and:

\begin{itemize}
    \item \(\big|0 \rangle \) represents strategy '0' of player \(i\). For example, in the game defined in Table \ref{tab:parameterPayoffMatrixMerging}, which addresses the use case illustrated in subfigure (b) of Figure \ref{fig:use_case}, it corresponds to the \textit{Merge} action for the ego vehicle. 
    \item \(\big|1 \rangle \) represents strategy '1' of player \(i\). In the same game setup (Table \ref{tab:parameterPayoffMatrixMerging} and subfigure (b) of Figure \ref{fig:use_case}), it corresponds to the \textit{Not Merge} action for player A (EV).
    \item \(\alpha_i\) is the complex probability amplitude of adopting strategy '0'.
    \item \(\beta_i\) is the complex probability amplitude of adopting strategy '1'. 
\end{itemize}

\begin{equation}\label{eq:|0>}
    \big|0 \rangle = 
    \begin{bmatrix}
        1 \\
        0
    \end{bmatrix}
\end{equation}

\begin{equation}\label{eq:|1>}
    \big|1 \rangle = 
    \begin{bmatrix}
        0 \\
        1
    \end{bmatrix}
\end{equation}

Furthermore, since \(\alpha_i\) and \(\beta_i\) are probability amplitudes, it implies:

\begin{equation} \label{eq:probability}
    \big|\alpha_i\big|^2 + \big|\beta_i\big|^2 = 1
\end{equation}

Thus, \(\big|\alpha_i\big|^2\) represents the probability of observing the outcome \(\big|0\rangle\), while \(\big|\beta_i\big|^2\) corresponds to the probability of observing the outcome \(\big|1\rangle\) for player \(i\). 

As illustrated in Figures \ref{fig::QG-U1} and \ref{fig::QG-G4}, \(\big| \psi_f \rangle\) is the final quantum state of each model. Concerning the QG-U1 model, \(\big| \psi_f \rangle\) is derived as follows:

\begin{equation} \label{eq:finalstate}
    \big|\psi_f \rangle= \mathcal{\hat{J}}^{\dagger} \cdot (\hat{U_a} \otimes \hat{U_b}) \cdot \mathcal{\hat{J}} \cdot \big|\psi_0\rangle 
\end{equation}

Where:
\begin{itemize}
    \item \(\big|\psi_0 \rangle\) is the initial quantum state of the system (Eq. \ref{eq:initialstate}).
    \item \(\hat{\mathcal{J}}\) is the entanglement operator (Eq. \ref{eq:J}), which introduces a connection (correlation) between the two players' choices in the quantum game. This correlation is controlled by the parameter \(\gamma\), which sets how strongly the players' decisions are linked. When \(\gamma =0\), the players' strategies are completely independent, just like in a classical game. As \(\gamma\) increases, their strategies become increasingly correlated. At the highest level of entanglement \(\gamma =\frac{\pi}{2}\), the players' strategies are so closely related that they can no longer be described separately (one player's decision instantly affects the other).    
    \item \(\hat{U}_i\) is the unitary operator of player \(i\). Basically, it is referred to as the quantum strategy of the player \(i\) (Eq. \ref{eq:unitary1parameters}).
    \item \(\mathcal{\hat{J}}^{\dagger}\) is the disentanglement operator, which allows the system to disentangle in order to perform measurements in the quantum circuit (Eq. \ref{eq:Jdagger}). 
    \item \(\big| \psi_f \rangle = [\psi_{f_{00}}, \psi_{f_{01}}, \psi_{f_{10}}, \psi_{f_{11}}]^T\) represents the final quantum state. Each element \(\psi_{f_{jk}}\) is a complex probability amplitude corresponding to the outcome when player A plays strategy \(j\) and player B plays strategy \(k\). 
\end{itemize}

\begin{equation} \label{eq:initialstate}
    \big| \psi_0 \rangle = q_0 \otimes q_1
\end{equation}

We defined \(\mathcal{\hat{J}}\) as follows:
\begin{equation} \label{eq:J}
            \mathcal{\hat{J}} = e^{-j\frac{\gamma}{2} \sigma_x \otimes \sigma_x}
\end{equation}

With \(0 \leq \gamma \leq \frac{\pi}{2}\) representing the entanglement factor (which describes how quantum the game is), and \(\sigma_x\) is the Pauli-X matrix (Eq. \ref{eq:Pauli}).

As for any matrix \(A\) where \(A^2=I\) (with \(I\) being the identity matrix), the matrix exponential simplifies via the Taylor series and results in the following:

\begin{equation} \label{eq:Jdevelopped}
    \mathcal{\hat{J}} = \begin{bmatrix}
        cos(\frac{\gamma}{2}) & 0 & 0 & -jsin(\frac{\gamma}{2})\\
        0 & cos(\frac{\gamma}{2}) & -jsin(\frac{\gamma}{2}) & 0\\
        0 & -jsin(\frac{\gamma}{2}) & cos(\frac{\gamma}{2}) & 0\\
        -jsin(\frac{\gamma}{2}) & 0 & 0 & cos(\frac{\gamma}{2})
        \end{bmatrix}
\end{equation}

Since \(\mathcal{\hat{J}}^{\dagger}\) being the Hermitian conjugate of \(\mathcal{\hat{J}}\), it involves that: 

\begin{equation}\label{eq:Jdagger}
    \mathcal{\hat{J}}^{\dagger} = \begin{bmatrix}
        cos(\frac{\gamma}{2}) & 0 & 0 & jsin(\frac{\gamma}{2})\\
        0 & cos(\frac{\gamma}{2}) & jsin(\frac{\gamma}{2}) & 0\\
        0 & jsin(\frac{\gamma}{2}) & cos(\frac{\gamma}{2}) & 0\\
        jsin(\frac{\gamma}{2}) & 0 & 0 & cos(\frac{\gamma}{2})
        \end{bmatrix}
\end{equation}

The unitary operator \(\hat{U_i}\) represents the quantum strategy of each player \(i\). As proposed in \cite{eisert1999quantum}, it is typically defined as follows:

\begin{equation} \label{eq:unitary2parameters}
    \hat{U_i} (\theta, \phi)=\begin{bmatrix}
            e^{j\phi}cos\frac{\theta}{2} & sin\frac{\theta}{2} \\
            -sin\frac{\theta}{2} & e^{-j\phi}cos\frac{\theta}{2} 
            \end{bmatrix}
\end{equation}

With \(0 \leq \theta \leq \pi\) affecting the level of superposition, and \(0 \leq \phi \leq \frac{\pi}{2}\) influencing the interference between qubits.

However, \(\hat{U_i}\) can also be defined using a three different parameters version, as in \cite{flitney2002introduction}, which allows for full control over the qubit state. For the sake of simplicity and to facilitate the analysis of the results, we assume \(\phi=0\). Thus, \(\hat{U_i}\) is defined as:

\begin{equation} \label{eq:unitary1parameters}
    \hat{U_i} (\theta)=\begin{bmatrix}
            cos\frac{\theta}{2} & sin\frac{\theta}{2} \\
            -sin\frac{\theta}{2} & cos\frac{\theta}{2} 
            \end{bmatrix}
\end{equation}

Later, when referring to this quantum model, we will refer to it as QG-U1 (for Quantum Game - Unitary 1).
As illustrated in Figure \ref{fig::QG-G4}, another solution to simplifying the unitary operator (Eq. \ref{eq:unitary2parameters}) and improving the analysis of its outcomes is to use quantum gates for each player \(i\), \(QG_i\), instead of the unitary operator \(\hat{U_i}\). Essentially, quantum gates are specific settings of the unitary operator that facilitate a more effective analysis of the outcome.

We define \(QG_i = \{H,\sigma_x,\sigma_y,\sigma_z,I_2\}\), where:

\begin{itemize}
    \item \(H\) is the Hadamard gate (Eq. \ref{eq:hadamard}),
    \item \(\sigma_x\), \(\sigma_y\), and \(\sigma_z\) are the Pauli gates (Eq. \ref{eq:Pauli}),
    \item \(I_2\) is the 2x2 identity matrix.
\end{itemize}

While other quantum gates can be employed, we have chosen to use these five specific quantum gates to make the visualization of the outcomes easier for the reader. Additional quantum gates can be found in \cite{aaronson2022introduction}.

\begin{equation} \label{eq:Pauli}
\begin{aligned}
    \sigma_x &= \begin{bmatrix}
        0 & 1 \\
        1 & 0
    \end{bmatrix}, \quad
    \sigma_y = \begin{bmatrix}
        0 & -j \\
        j & 0
    \end{bmatrix}, \quad
    \sigma_z = \begin{bmatrix}
        1 & 0 \\
        0 & -1
    \end{bmatrix}
\end{aligned}
\end{equation}

\begin{equation} \label{eq:hadamard}
    H = \frac{1}{\sqrt2} \begin{bmatrix}
        1 & 1 \\
        1 & -1
    \end{bmatrix}
\end{equation}

Later, when referring to this quantum game model, we will refer to it as QG-G4 (Quantum Game - Gates 4).

\section{Results}\label{sec::results}
\begin{figure}[h]
\centering
\subfigure[Roundabout]
{\includegraphics[width=.8\linewidth]{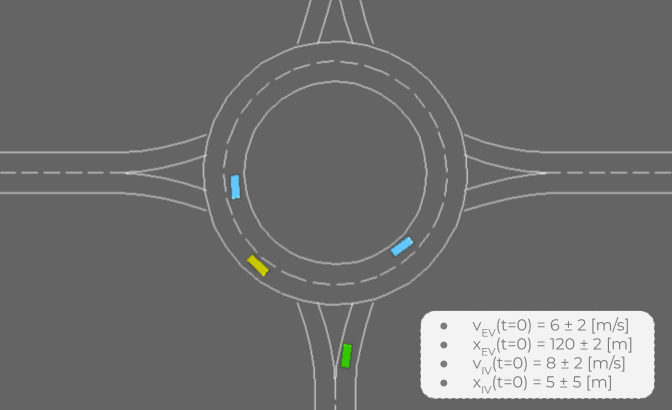}}\\
\subfigure[Merging]{\includegraphics[width=.9\linewidth]{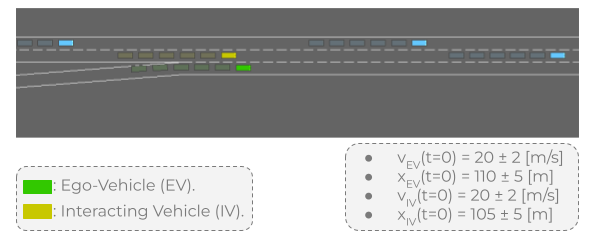}}
\caption{The two scenarios as presented in this paper, with the images generated using the highway-env simulator \cite{highway-env}.}
\label{fig:use_case}
\end{figure}

To evaluate our quantum models in a driving scenario, we test them in situations involving strong interactions between road users, such as merging and roundabouts, which are both depicted in Figure \ref{fig:use_case}. These two use cases are situations where coordination, cooperation, and synergy are required for each involved agent. To define these two scenarios as a game, we consider two agents: the ego vehicle (EV) and an Interacting Vehicle (IV). In each game, each agent has two pure strategies. For the merging scenario: \(\mathcal{A}_{EV}=\{\text{Merge, Not\ Merge}\}\), and \(\mathcal{A}_{IV}=\{\text{Accelerate, Decelerate}\}\). Concerning the roundabout, we consider this set of actions: \(\mathcal{A}_{EV}=\{\text{Accelerate, Decelerate}\}\), and \(\mathcal{A}_{IV}=\{\text{Accelerate, Idle}\}\). This game setting results in four possible states, denoted \(s_{jk} \in \mathcal{S}\), where \(j\) is the action index of the EV and \(k\) the action index of the IV. The utilities associated with each state, along with the defined pure strategies, are summarized in Tables \ref{tab:parameterPayoffMatrixRoundabout} and  \ref{tab:parameterPayoffMatrixMerging}. These two payoff matrices each contain two Nash Equilibria: N.E.(1), corresponding to states \(s_{10}\), and N.E.(2), corresponding to state \(s_{01}\), in both games. Since both equilibria lie on the diagonal of the payoff matrix, these two games present a dilemma and cannot be resolved through standard equilibrium selection techniques. In such cases, probabilistic methods are commonly employed to address the dilemma. 

\begin{table}[ht]
    \centering
\begin{tabular}{ccc|c}
    & & \multicolumn{2}{c}{\textbf{Player 2 (IV)}} \\
    \cline{3-4}
    &  & Accelerate (A) & Idle (I) \\
    \hline
    \multirow[c]{2}{*}{\textbf{Player 1 (EV)}} & Accelerate (A) & 0, 0 & 10, 4\\
    \cline{2-4}
    & Decelerate (D)& 4, 10 & 4, 4\\
    \hline
\end{tabular}
    \caption{Game Parameters - Utilities and pure strategies used in the roundabout scenario depicted in subfigure (a) of Figure \ref{fig:use_case}.}
    \label{tab:parameterPayoffMatrixRoundabout}
\end{table}

\begin{table}[h!]
    \centering
\begin{tabular}{ccc|c}
    & & \multicolumn{2}{c}{\textbf{Player 2 (IV)}} \\
    \cline{3-4}
    &  & Accelerate (A) & Decelerate (D) \\
    \hline
    \multirow[c]{2}{*}{\textbf{Player 1 (EV)}} & Merge (M) & 0, 0 & 10, 4\\
    \cline{2-4}
    & Not\ Merge (N)& 4, 10 & 1, 1\\
    \hline
\end{tabular}
    \caption{Game Parameters - Utilities and pure strategies used in the merging scenario depicted in subfigure (b) of Figure \ref{fig:use_case}.}
    \label{tab:parameterPayoffMatrixMerging}
\end{table}

\begin{figure}[!ht]
    \centering
    \includegraphics[width=0.95\linewidth]{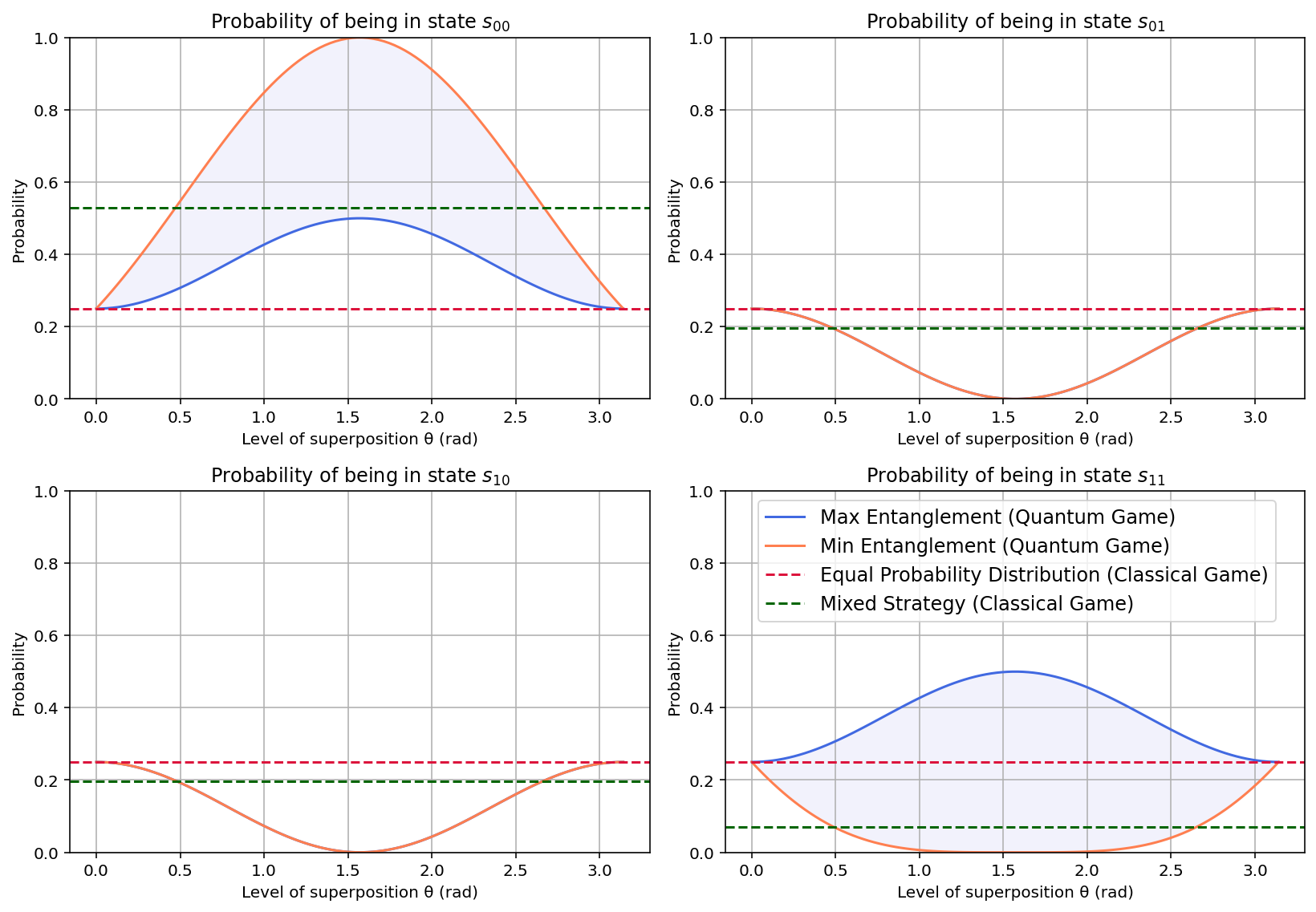}
    \caption{Probabilities of being in each state of the merging game obtained from different models. For the quantum game, the probabilities result from the QG-U1 model while varying the level of superposition \(\theta\), with the initial state \(\big| \psi_0 \big|^2 = [0.25,0.25,0.25,0.25]^T\).}
    \label{fig:probabilities-QG-U1}
\end{figure}

As mentioned in the introduction, one limitation of classical game theory lies in its predictability - once the strategies and utilities are known, the outcome becomes relatively deterministic. In contrast, quantum games introduce a layer of unpredictability due to quantum effects, making it harder for opponents to predict the outcome. Figure \ref{fig:probabilities-QG-U1} illustrates this phenomenon. It shows the probability of being in each state in the merging game for classical and quantum games. The red and green dashed plots represent values obtained using classical game theory, where probabilities remain fixed. The green plot is obtained using the mixed strategy technique (Equations \ref{eq::mixedstrategyIV} and \ref{eq::mixedstrategyEV}), while the red plot corresponds to an equal probability distribution. These constant probability values contribute to the predictability observed in classical games.
In contrast, the quantum game (QG-U1 in this case) demonstrates dynamic behavior: the probabilities of being in each state vary with quantum parameters, particularly \(\theta\) and \(\gamma\). For clarity, this figure assumes \(\theta_{EV}=\theta_{IV}\). The blue plot corresponds to a fully entangled quantum setting: \(\gamma=\frac{\pi}{2}\), while the orange plot represents a non-entangled setting: \(\gamma=0\). The surface between them illustrates the probabilities as \(\gamma\) varies between these extremes.
These results were obtained using an initial quantum state following an equal probability distribution. However, we observed that modifying the initial state leads to entirely different probability surfaces.

As described in the section outlining our approach, several variables define a quantum game (such as \(\gamma,\ \theta, \ \phi\)). These parameters directly influence the quantum probabilities and, consequently, the outcome of the model. So, the first step is to look over the impact of these different parameters to understand the model's behavior and possibly cluster some models from it. To achieve this, we analyzed the expected utility of the EV (Player A) as a function of the quantum parameters. As discussed by Schoemaker in \cite{schoemaker1982expected}, in his expected utility model, people under uncertainty tend to make decisions that maximize their expected utility. In a game, the expected utility of a player \(i\), denoted \(\mathbb{E}(u_i)\) (Eq. \ref{eq::expected_utility}), is computed based on a utility function \(u_i\), which quantifies the desirability of each outcome, and a probability \(p\), which captures the likelihood of that outcome occurring. In this paper, we also refer to the expected utility as expected payoff, as both terms are equivalent in our context.

\begin{equation} \label{eq::expected_utility}
    \mathbb{E}(u_i)=\sum_{a \in \mathcal{A}} p(a)\cdot u_i(a)
\end{equation}

\begin{figure*}[h]
\subfigure[QG-U1]{\includegraphics[width=.33\textwidth]{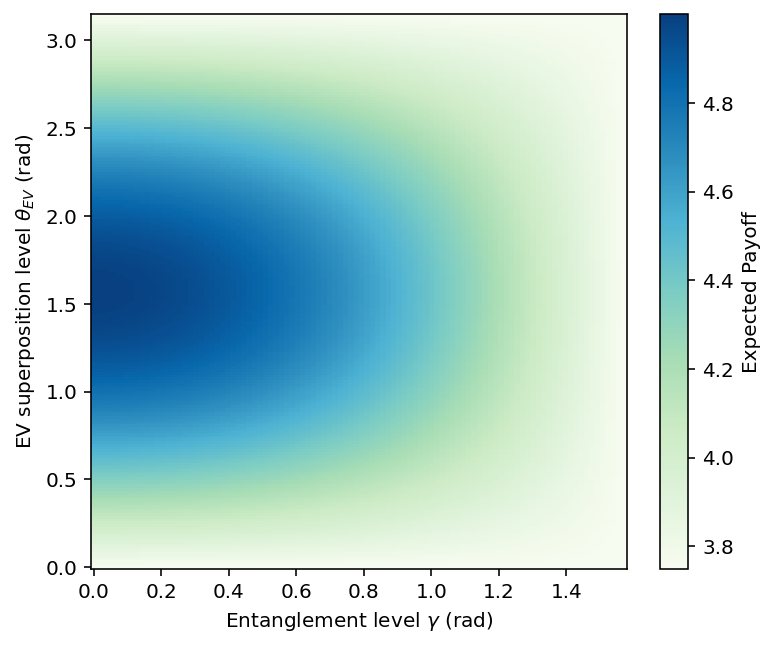}}\hfill
\subfigure[QG-G4 - \(\gamma_{max}\)]{\includegraphics[width=.33\textwidth]{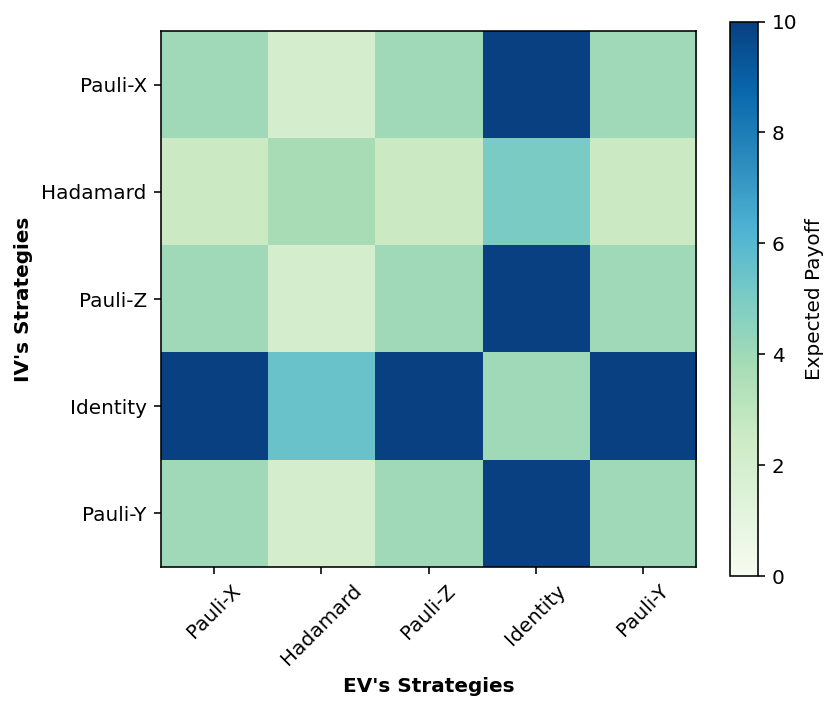}}\hfill 
\subfigure[QG-G4 - \(\gamma_{min}\)]{\includegraphics[width=.33\textwidth]{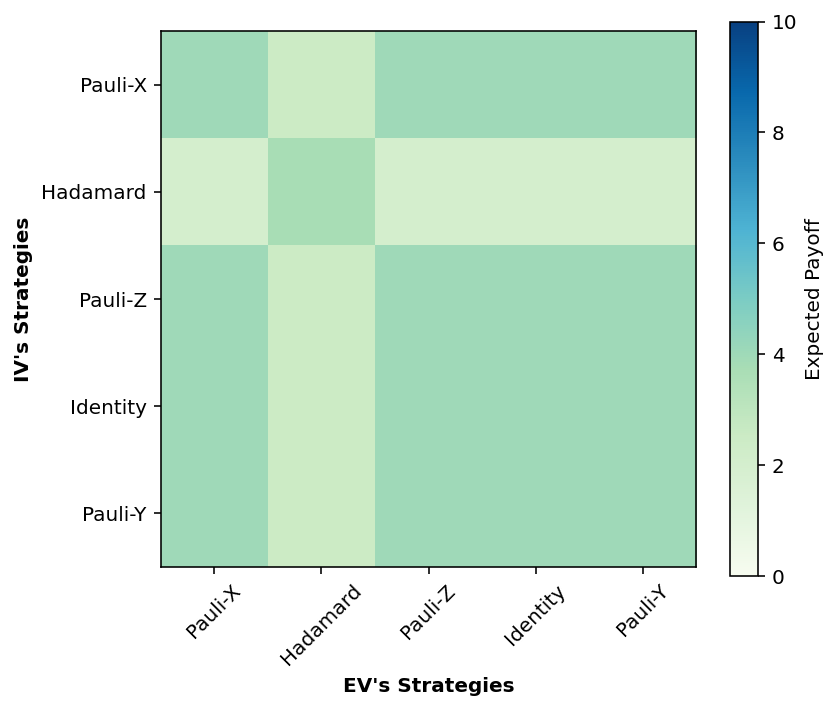}}\hfill
\caption{Subfigure (a): Expected payoff obtained from QG-U1 model (Figure \ref{fig::QG-U1}) while \(\theta_{IV}=0\), and \(\big| \psi_0 \big|^2 = [0.25,0.25,0.25,0.25]^T\). Subfigures (b) and (c): Expected payoff obtained from QG-G4 model (Figure \ref{fig::QG-G4}) when \(\big| \psi_0 \big|^2 = [0,0,1,0]^T\) for different quantum gates played. Subfigure (b): results obtained with \(\gamma=\frac{\pi}{2}\), meaning a maximum level of entanglement. Subfigure (c): results obtained with \(\gamma=0\), meaning a minimum level of entanglement.}
\label{fig:expected_payoff}
\end{figure*}

The subfigures (a), (b), and (c) of Figure \ref{fig:expected_payoff} are obtained with utility values from the merging game. Subfigure (a) of Figure \ref{fig:expected_payoff} illustrates the evolution of the expected payoff for the EV as a function of \(\gamma\) and \(\theta_{EV}\), using the QG-U1 model. In this subfigure, we assume that \(\theta_{IV}=0\), meaning that the interacting vehicle plays a classical game instead of a quantum one. This assumption also allows more clarity in the figure, since one dimension is not considered. Furthermore, the input quantum state is assumed to follow an equal probability distribution. From this figure, we extract two specific models:
\begin{itemize}
    \item \textbf{QG-U1-1}, corresponding to the configuration that maximizes \(\mathbb{E}(u_{EV})\): \(\gamma=0\), \(\theta_{EV}=\frac{\pi}{2}\), and \(\theta_{IV}=0\).
    
    \item \textbf{QG-U1-2}, corresponding to the configuration that minimizes \(\mathbb{E}(u_{EV})\): \(\gamma=
    \frac{\pi}{2}\), \(\theta_{EV}=0\), and \(\theta_{IV}=0\).
\end{itemize}

Subfigures (b) and (c) of Figure \ref{fig:expected_payoff} show the expected payoff for the EV when using the QG-G4 model. Subfigure (b) corresponds to the case of maximum entanglement (\(\gamma=\frac{\pi}{2}\)), while subfigure (c) represents the case of minimum entanglement (\(\gamma=0\)). 
Concerning this model, we observed that, when the input follows an equal probability distribution, the different quantum gates \(QG\) do not have a significant influence on the model's outcome, except for the Hadamard gate \(H\), due to its ability to place the system in a superposition state. However, when the input does not follow an equal probability distribution, the outcomes change considerably. Specifically, subfigures (b) and (c) were obtained when \(\big| \psi_0 \big|^2 = [0,0,1,0]^T\). This means that, at the input of the model, the system is in the state \(s_{10}\). Furthermore, we noticed that even though the Pauli matrices \(\sigma_x\) and \(\sigma_y\) are different, they yield the same outcomes.
Another notable observation is that certain configurations yield an expected payoff of \(\mathbb{E}(u_{EV}) = 10\), which is almost impossible to reach in a classical game. For example, when \(\gamma=\frac{\pi}{2}\), and the EV plays "\(I_2\)" while the IV plays "\(\sigma_z\)", the EV reaches this optimal expected utility.
Here, we can observe that the expected payoff is maximized when the EV plays the identity gate \(I_2\) and the entanglement parameter is set to \(\gamma=\frac{\pi}{2}\). Therefore, when referring to the QG-G4 model later in this paper, we adopt this configuration: the EV always plays \(I_2\) and the entanglement level is set to \(\gamma=\frac{\pi}{2}\).

\begin{figure}
    \centering
    \includegraphics[width=\linewidth]{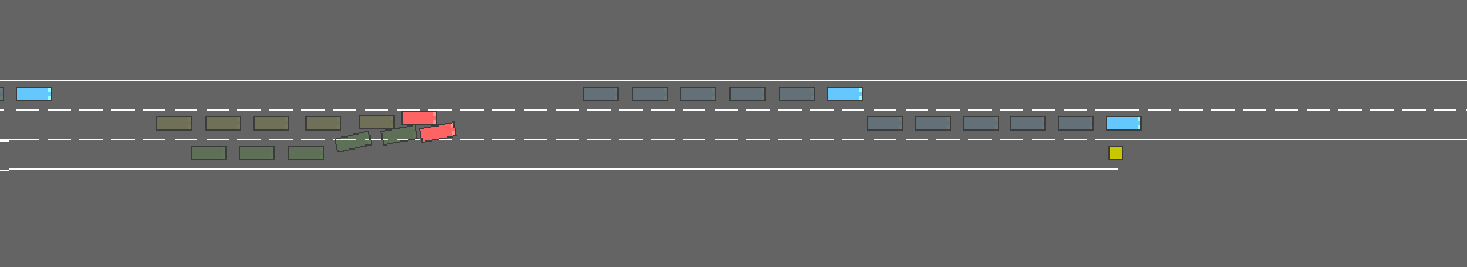}
    \caption{State \(s_{00}\) in the merging game which leads to a collision: the ego vehicle decides to merge and the interacting vehicle chooses to accelerate.}
    \label{fig:highway-env-00}
\end{figure}

\begin{figure}
    \centering
    \includegraphics[width=\linewidth]{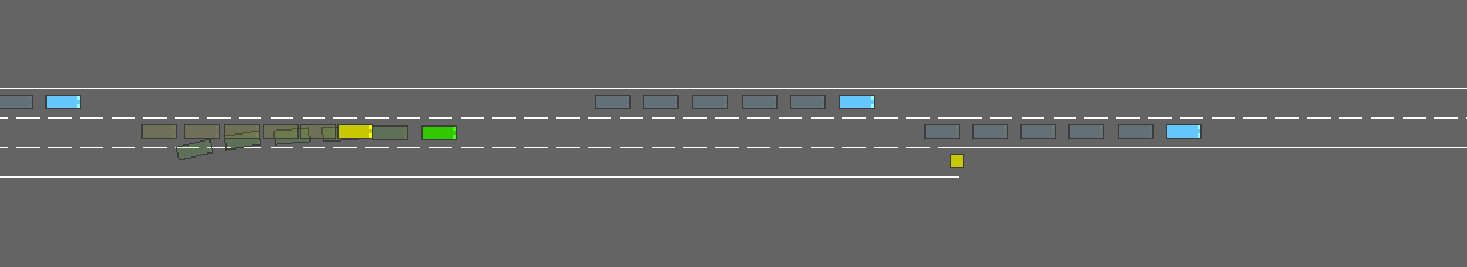}
    \caption{State \(s_{01}\) in the merging game which leads to a successful merging: the ego vehicle merges, and the interacting vehicle decelerates to let the EV merge.}
    \label{fig:highway-env-01}
\end{figure}

\begin{table}[b!]
    \centering
    \begin{tabular}{ccccc}
    \hline
    Scenario & Method & CR $\downarrow$ & SR $\uparrow$ & \(\bar{HD}\) $\uparrow$\\
    \hline
    \multirow{7}{*}{Merging} & COR-MP \cite{cor-mp} & 3.5\% & 84.1\% & - \\
    & MOBIL \cite{MOBIL} & 3.1\% & 83.9\% & - \\
    \cline{2-5}
    \addlinespace
    & CG-EPD & 25.23\% & 50.19\% & - \\
    & CG-MS & 48.01\% & 43.02\% & - \\
    \cline{2-5}
    \addlinespace
    & \textbf{QG-G4} & \textbf{2.8\%} & \textbf{90.15\%} & -\\
    & QG-U1-1 & 50.07\% & 49.93\% & - \\
    & QG-U1-2 & 25.29\% & 49.54\% & - \\
    \hline
    \addlinespace
    \multirow{7}{*}{Roundabout} & COR-MP \cite{cor-mp} & 20.5\% & 79.5\% & 7.13m \\
    & IDM \cite{IDM} & 7.9\% & 92.1\% & 12.14m \\
    \cline{2-5}
    \addlinespace
    & CG-EPD & 47.5\% & 52.5\% & 8.28m\\
    & CG-MS & 28.6\% & 71.4\% & 6.02m\\
    \cline{2-5}
    \addlinespace
    & \textbf{QG-G4} & \textbf{1.3\%} & \textbf{98.7\%} & \textbf{12.53m} \\
    & QG-U1-1 & 19\% & 81\% & 5.06m\\
    & QG-U1-2 & 33.2\% & 66.8\% & 4.79m\\
    \hline
    
    \end{tabular}
    \caption{Quantitative results obtained for the two scenarios presented in this paper (Figure \ref{fig:use_case}). Compared with baseline methods. \textbf{CR}: Collision rate, \textbf{SR}: Success rate, and \(\mathbf{\bar{HD}}\): Mean Headway Distance (between EV and IV).}
    \label{tab:quantitativeresults}
\end{table}

Finally, to evaluate the quantum models, we run both use cases thousands of times using the highway-env simulator \cite{highway-env}, varying the initial positions and speeds of the EV and IV in each episode, as specified in the legend of Figure \ref{fig:use_case}. As seen above, since the two N.E. are located on the diagonal of the payoff matrix, solving this game is challenging. One classical approach to address this is through probabilistic modeling. For the classical game, we use two models:

\begin{itemize}
    \item \textbf{CG-EPD} (Classical Game - Equal Probability Distribution): This model assumes a uniform probability distribution over all strategies.

    \item \textbf{CG-MS} (Classical Game - Mixed Strategy): In this model, each pure strategy is assigned a probability based on the players' utilities. With \(p\) being the probability that the IV chooses the strategy \(a_1\) and \(1-p\) the probability it chooses \(a_2\). Likewise, \(q\) is the probability that the EV chooses \(a_1\), and \(1-q\) is the probability it chooses \(a_2\). Equations \ref{eq::mixedstrategyIV} and \ref{eq::mixedstrategyEV} detail the computation of these probabilities.
\end{itemize}

\begin{equation}
    \label{eq::mixedstrategyIV}
    p = \frac{u^{EV}_{11} - u^{EV}_{01}}{u^{EV}_{00} - u^{EV}_{01} - u^{EV}_{10} + u^{EV}_{11}}
\end{equation}

\begin{equation}
    \label{eq::mixedstrategyEV}
    q =\frac{u^{IV}_{11} - u^{IV}_{10}}{u^{IV}_{00} - u^{IV}_{01} - u^{IV}_{10} + u^{IV}_{11}}
\end{equation}

For both quantum and classical game models, the final ego decision is made probabilistically, following the probability output by each model. For the IV decision, to make it behave uncertain, its decision is made following an equal probability distribution; it chooses \(a_1\) 50\% of the time, and the other 50\% it chooses \(a_2\).

Table \ref{tab:quantitativeresults} highlights the quantitative results obtained for the two use cases presented in this paper. Since we run several times each use case, the values correspond to the distribution and the mean. In this table, we include three parameters: \textbf{CR} representing the Collision Rate, \textbf{SR} representing the Success Rate, whereas the EV success to merge to the left lane or if it success to enter the roundabout without any collision and with no violation of the traffic rules, and finally \(\mathbf{\bar{HD}}\) which represents the Mean Headway Distance between the EV and IV. From this table, we made some observations. First, regarding the classical game models, we observed that the CG-MS model often ends up in state \(s_{00}\), which corresponds to a collision (as shown in Figure \ref{fig:highway-env-00}). This outcome is expected, as the probabilities in this model are computed based on utilities - each agent tends to maximize its own payoff. For the quantum models, we noted the following observations: 

\begin{itemize}
    \item \textbf{QG-U1-1}: This model most frequently results in states \(s_{00}\) and \(s_{01}\). Indeed, under this configuration, the EV consistently chooses to merge (in the merging game) and to accelerate (in the roundabout game). It appears that the action \(a_1\) becomes dominant for the EV.
    
    \item \textbf{QG-U1-2}: The distribution among the states observed here is similar to that of the CG-EPD model, suggesting behavior close to a uniform probability distribution.

    \item \textbf{QG-G4}: Two notable results emerge from this model. First, the collision state \(s_{00}\) occurs only 1.3\% of the time in the roundabout use case and 2.8\% in the merging scenario, making QG-G4 the model with the fewest collisions among those tested. It also achieves the highest success rate (\textbf{SR}) while maintaining a low collision rate (\textbf{CR}). QG-G4 demonstrates strong adaptability to the IV's uncertain behavior and reaches Nash Equilibrium more often than other models.
\end{itemize}


We also compared with other baseline methods. The first is the Conservation of Resources model for Maneuver Planning (COR-MP) \cite{cor-mp}, a utility-based maneuver planner. For the merging scenario, we also include the Minimizing Overall Braking Induced by Lane Changes (MOBIL) model \cite{MOBIL}, and for the roundabout use case, the Intelligent Driver Model (IDM) \cite{IDM}. Both MOBIL and IDM are configured with a regular driver profile.
We noticed that COR-MP, most of the time, adopts a conservative behavior. In the merging scenario, for instance, it often chooses to decelerate and merge after the IV passes. However, because the IV follows an uncertain policy, this sometimes leads to both vehicles traveling next to each other along the merging ramp, occasionally causing the EV to become stuck and unable to merge.

\section{Conclusion} \label{sec::conclusion}

In this paper, we present QG-U1 and QG-G4, two quantum models designed to address interaction-aware decision-making in automated driving. These models combine classical game theory with principles of quantum mechanics, such as superposition, interference, and entanglement, yielding outcomes that differ from classical game approaches, and, in some cases, provide higher expected payoffs.
Our results show that quantum games offer promising outcomes in scenarios involving high interactions where traditional methods tend to favor conservative behavior. Moreover, QG-U1 and QG-G4 are adaptable and can be extended to consider other actions depending on the actual driving context. We believe integrating quantum models into automated driving decision-making can lead to promising outcomes, particularly in environments with complex interactions. Nevertheless, further research is required to fully assess its potential. Future work will focus on applying these models to broader driving scenarios (e.g., highways and intersections) and designing a payoff matrix that adapts in real time to the environment. 

\bibliographystyle{IEEEtran}
\bibliography{bibliography}

\begin{thebibliography}{10}
\providecommand{\url}[1]{#1}
\csname url@samestyle\endcsname
\providecommand{\newblock}{\relax}
\providecommand{\bibinfo}[2]{#2}
\providecommand{\BIBentrySTDinterwordspacing}{\spaceskip=0pt\relax}
\providecommand{\BIBentryALTinterwordstretchfactor}{4}
\providecommand{\BIBentryALTinterwordspacing}{\spaceskip=\fontdimen2\font plus
\BIBentryALTinterwordstretchfactor\fontdimen3\font minus \fontdimen4\font\relax}
\providecommand{\BIBforeignlanguage}[2]{{%
\expandafter\ifx\csname l@#1\endcsname\relax
\typeout{** WARNING: IEEEtran.bst: No hyphenation pattern has been}%
\typeout{** loaded for the language `#1'. Using the pattern for}%
\typeout{** the default language instead.}%
\else
\language=\csname l@#1\endcsname
\fi
#2}}
\providecommand{\BIBdecl}{\relax}
\BIBdecl

\bibitem{advancements}
L.~Haj~Meftah and A.~Cherif, ``{Exploring Autonomous Vehicle Technology: Advancements, Challenges, and the Critical Role of Simulation},'' \emph{Asma, Exploring Autonomous Vehicle Technology: Advancements, Challenges, and the Critical Role of Simulation}.

\bibitem{cor-mp}
K.~Essalmi, F.~Garrido, and F.~Nashashibi, ``{COR-MP: Conservation of Resources Model for Maneuver Planning},'' in \emph{2024 IEEE 20th International Conference on Intelligent Computer Communication and Processing (ICCP)}.\hskip 1em plus 0.5em minus 0.4em\relax IEEE, 2024, pp. 1--8.

\bibitem{cor-mcts}
------, ``{An Extended Horizon Tactical Decision-Making for Automated Driving Based on Monte Carlo Tree Search},'' in \emph{IV 2025-36th IEEE Intelligent Vehicles Symposium}, 2025.

\bibitem{von2007theory}
J.~Von~Neumann and O.~Morgenstern, ``{Theory of games and economic behavior: 60th anniversary commemorative edition},'' in \emph{Theory of games and economic behavior}.\hskip 1em plus 0.5em minus 0.4em\relax Princeton university press, 2007.

\bibitem{che2024game}
C.~Che and J.~Tian, ``{Game theory: Concepts, applications, and insights from operations research},'' \emph{Journal of Computer Technology and Applied Mathematics}, vol.~1, no.~4, pp. 53--59, 2024.

\bibitem{songvehicle}
Q.~Song, W.~Fu, W.~Wang, Y.~Sun, D.~Wang, and J.~Zhou, ``{Quantum decision making in automatic driving},'' \emph{Scientific reports}, 2022.

\bibitem{silva2022learning}
A.~Silva, O.~G. Zabaleta, and C.~M. Arizmendi, ``{Learning mixed strategies in quantum games with imperfect information},'' \emph{Quantum Reports}, vol.~4, no.~4, pp. 462--475, 2022.

\bibitem{zhang2021subjective}
C.~Zhang and H.~Kjellstr{\"o}m, ``{A subjective model of human decision making based on quantum decision theory},'' \emph{arXiv preprint arXiv:2101.05851}, 2021.

\bibitem{songpedestrian}
Q.~Song, W.~Wang, W.~Fu, Y.~Sun, D.~Wang, and Z.~Gao, ``{Research on quantum cognition in autonomous driving},'' \emph{Scientific reports}, 2022.

\bibitem{sinha2025nav}
A.~Sinha, A.~Macaluso, and M.~Klusch, ``{Nav-Q: quantum deep reinforcement learning for collision-free navigation of self-driving cars},'' \emph{Quantum Machine Intelligence}, vol.~7, no.~1, pp. 1--20, 2025.

\bibitem{garrido2022review}
F.~Garrido and P.~Resende, ``{Review of decision-making and planning approaches in automated driving},'' \emph{IEEE Access}, 2022.

\bibitem{ibrahim2021comprehensive}
M.~A.~R. Ibrahim, N.~I. Jaini, and K.~M. N.~K. Khalif, ``{A comprehensive review of hybrid game theory techniques and multi-criteria decision-making methods},'' in \emph{Journal of Physics: Conference Series}, 2021.

\bibitem{naidja2024gtp}
N.~Naidja, M.~Revilloud, S.~Font, and G.~Sandou, ``{GTP-UDrive: Unified Game-Theoretic Trajectory Planner and Decision-Maker for Autonomous Driving in Mixed Traffic Environments},'' in \emph{2024 IEEE Intelligent Vehicles Symposium (IV)}.\hskip 1em plus 0.5em minus 0.4em\relax IEEE, 2024, pp. 3262--3268.

\bibitem{liu2022three}
M.~Liu, Y.~Wan, F.~L. Lewis, S.~Nageshrao, and D.~Filev, ``{A three-level game-theoretic decision-making framework for autonomous vehicles},'' \emph{IEEE Transactions on Intelligent Transportation Systems}, 2022.

\bibitem{burger2022interaction}
C.~Burger, J.~Fischer, F.~Bieder, {\"O}.~{\c{S}}. Ta{\c{s}}, and C.~Stiller, ``{Interaction-aware game-theoretic motion planning for automated vehicles using bi-level optimization},'' in \emph{2022 IEEE 25th International Conference on Intelligent Transportation Systems (ITSC)}.\hskip 1em plus 0.5em minus 0.4em\relax IEEE, 2022.

\bibitem{garzon2019game}
M.~Garz{\'o}n and A.~Spalanzani, ``{Game theoretic decision making for autonomous vehicles’ merge manoeuvre in high traffic scenarios},'' in \emph{2019 IEEE Intelligent Transportation Systems Conference (ITSC)}.\hskip 1em plus 0.5em minus 0.4em\relax IEEE, 2019, pp. 3448--3453.

\bibitem{yuan2023scalable}
M.~Yuan, J.~Shan, and H.~Schofield, ``{Scalable game-theoretic decision-making for self-driving cars at unsignalized intersections},'' \emph{IEEE Transactions on Industrial Electronics}, 2023.

\bibitem{heshami2024towards}
S.~Heshami and L.~Kattan, ``{Towards Self-Organizing connected and autonomous Vehicles: A coalitional game theory approach for cooperative Lane-Changing decisions},'' \emph{Transportation Research Part C: Emerging Technologies}, vol. 166, p. 104789, 2024.

\bibitem{huang2024integrated}
Z.~Huang, T.~Li, S.~Shen, and J.~Ma, ``{Integrated Decision Making and Trajectory Planning for Autonomous Driving Under Multimodal Uncertainties: A Bayesian Game Approach},'' \emph{arXiv preprint arXiv:2409.13993}, 2024.

\bibitem{zhou2024game}
X.~Zhou, Z.~Peng, Y.~Xie, M.~Liu, and J.~Ma, ``{Game-Theoretic Driver Modeling and Decision-Making for Autonomous Driving with Temporal-Spatial Attention-Based Deep Q-Learning},'' \emph{IEEE Transactions on Intelligent Vehicles}, 2024.

\bibitem{yuan2021deep}
M.~Yuan, J.~Shan, and K.~Mi, ``{Deep reinforcement learning based game-theoretic decision-making for autonomous vehicles},'' \emph{IEEE Robotics and Automation Letters}, vol.~7, no.~2, pp. 818--825, 2021.

\bibitem{pricequantum}
E.~Price, ``{Quantum Games and Game Strategy}.''

\bibitem{wiki_quantum_game_theory}
\BIBentryALTinterwordspacing
{Wikipedia contributors}, ``Quantum game theory.'' [Online]. Available: \url{https://en.wikipedia.org/wiki/Quantum\_game\_theory}
\BIBentrySTDinterwordspacing

\bibitem{eisert1999quantum}
J.~Eisert, M.~Wilkens, and M.~Lewenstein, ``{Quantum games and quantum strategies},'' \emph{Physical Review Letters}, 1999.

\bibitem{flitney2002introduction}
A.~P. Flitney and D.~Abbott, ``{An introduction to quantum game theory},'' \emph{Fluctuation and Noise Letters}, 2002.

\bibitem{khan2025quantum}
F.~S. Khan, N.~M. Linke, A.~T. Than, and D.~Baron, ``{Quantum Advantage in Trading: A Game-Theoretic Approach},'' 2025.

\bibitem{pothos2009quantum}
E.~M. Pothos and J.~R. Busemeyer, ``{A quantum probability explanation for violations of ‘rational’decision theory},'' \emph{Proceedings of the Royal Society B: Biological Sciences}, 2009.

\bibitem{aaronson2022introduction}
S.~Aaronson, ``{Introduction to quantum information science II lecture notes},'' 2022.

\bibitem{highway-env}
E.~Leurent, ``{An Environment for Autonomous Driving Decision-Making},'' \url{https://github.com/eleurent/highway-env}, 2018.

\bibitem{schoemaker1982expected}
P.~J. Schoemaker, ``{The expected utility model: Its variants, purposes, evidence and limitations},'' \emph{Journal of economic literature}, 1982.

\bibitem{MOBIL}
A.~Kesting, M.~Treiber, and D.~Helbing, ``{General lane-changing model MOBIL for car-following models},'' \emph{Transportation Research Record}, vol. 1999, 2007.

\bibitem{IDM}
M.~Treiber, A.~Hennecke, and D.~Helbing, ``{Congested traffic states in empirical observations and microscopic simulations},'' \emph{Physical review E}, 2000.

\end{thebibliography}
\end{document}